\documentclass[12pt,letterpaper]{JHEP3}
\usepackage{amsmath, amssymb, mathtools, stmaryrd}
\usepackage{amsthm}
\usepackage{amsfonts}
\usepackage{epsfig,multicol,float}

\def\be{\begin{equation}}
\def\ee{\end{equation}}
\def\bea{\begin{equation}\begin{aligned}}
\def\eea{\end{aligned}\end{equation}}
\def\be{\begin{eqnarray}}
\def\ee{\end{eqnarray}}
\def\beq{\begin{equation}}
\def\eeq{\end{equation}}

\def\({\left (}
\def\){\right )}
\def\[{\left [}
\def\[{\right ]}

\newcommand{\cO}{\mathcal{O}}
\title{Analytical Computation of Critical Exponents in Several Holographic Superconductors}

\author{Hua-Bi Zeng$^{1}$, Xin Gao$^{2}$, Yu Jiang$^{3}$ and Hong-Shi Zong$^{1,4}$\\

$^{1}$ Department of Physics, Nanjing University, Nanjing 210093, China\\

$^{2}$ Key Laboratory of Frontiers in Theoretical Physics, Institute of Theoretical
Physics, Chinese Academy of Sciences, Beijing 100190, China\\

$^{3}$ Center for Statistical and Theoretical Condensed Matter Physics \& Department of Physics, Zhejiang Normal University,  Jinhua 321004, China\\

$^{4}$ Joint Center for Particle, Nuclear Physics and Cosmology, Nanjing University, Nanjing 210093, China}

\abstract{
It is very interesting that all holographic superconductors, such as $s$-wave, $p$-wave and $d$-wave holographic superconductors, show the universal mean-field critical exponent $1/2$ at the critical temperature, just like Gindzburg-Landau (G-L) theory for second order phase transitions. Now it is believed that the universal
critical exponents appear because the dual gravity theory is classic in the large $N$ limit.
However, even in the large $N$ limit there is an exception called ``non-mean-field theory'': an extension of the $s$-wave model with a cubic term of the charged scalar field shows a different critical exponent $1$. In this paper, we try to use analytical methods to obtain the
critical exponents for these models to see how the properties of the
gravity action decides the appearance of the mean-field behaviors. It will be seen that just like the G-L theory, it is the fundamental symmetries rather than the detailed parameters of the bulk theory that lead to the universal properties of the holographic superconducting phase transition.
The feasibility of the called ``non-mean-field theory'' is also discussed.}

\begin{document}
\section{ Introduction}
Using AdS/CFT correspondence\cite{1,2,3,4} to study superconductors is
a very innovative idea. The method to build the holographic superconductors that we use here is called the ``bottom-up'' approach: by putting some set of fields in the asymptotic AdS spacetime, through the study of the weak interacting bulk field theory we can get the information of the strongly coupled boundary theory. For example, when putting a charged scalar field coupled to a $U(1)$ gauge field in the bulk, we get the $s$-wave holographic superconductor in which the order parameter is a scalar\cite{5,6}. By putting a pure $SU(2)$ gauge field in the bulk, we can get the $p$-wave holographic superconductor with a vector order
parameter \cite{7,8,9}. In order to build a $d$-wave holographic
superconductor we need a charged tensor field coupled to a $U(1)$ gauge
field in the bulk that leads to a tensor order parameter \cite{10,11,12}.
Besides the ``bottom-up'' approach, there is also a``top-down'' way to embed the holographic superconductors into string theory, since the scalar, gauge and tensor fields we used above have a natural source in string theory. In Ref. \cite{13} the authors studied various D-brane configurations for the $p+ip$-wave and $s$-wave model, whereas the embedding of the $d$-wave model with tensor field into string theory is still an open question \cite{12}.

These holographic superconductors have many properties similar to
real world superconductors: the second order phase transition at the critical temperature, the behaviors of holographic superconductor under a magnetic field, the AC conductivity, and so on.
One of the most important success of holographic superconductors is that all these models show the exact mean-field behaviors at the critical temperature, just like the Gindzburg-Landau (G-L) theory for continuous phase transitions. All the critical exponents
for the order parameter at $T_c$ are $1/2$. The G-L theory is a tremendous breakthrough
towards understanding the physical spirit of the continuous phase transition which enables us to understand that it is the fundamental symmetry that decides
the universality of continuous phase transitions. Till now, the study of holographic
superconductors made us believe that the AdS/CFT correspondence indeed contain the physics of real world superconductors. We know that the fundamental symmetry
decides the rule of the G-L theory. It is also a meaningful thing to answer whether this also works in the holographic models. However, most of the computations to
obtain the critical exponents are numerical. It is
hard to see exactly what properties of the action decide the
mean-filed behaviors. In this paper, we employ the analytical
method \cite{14} to stuty the $s$-wave, $p$-wave, $d$-wave and extended $s$-wave
holographic superconductors. There are other papers that study holographic
superconductors with analytical methods\cite{34,26}.
 The analytical calculation indeed provides more information to answer where the universal critical exponents $1/2$ of the order parameter comes from. It is
indeed the fundamental symmetry of the bulk theory that leads to the expected mean-field
behaviors.

The organization of this paper is as follows. In Section
$2$ we analytically compute the critical exponents at $T_c$
for $s$-wave, $d$-wave and $p$-wave models, respectively.
We can see how the fundamental symmetry of the actions decide the
critical exponents. In Section $3$ we give the computation of the
extended $s$-wave model with mean-field behavior and give some comments on the critical
exponents different from $1/2$ in the extended $s$-wave model. Discussions and conclusions are given in Section $4$.

\section{The $s$-wave model, $p$-wave model and $d$-wave model }
The construction of holographic superconductors is based on the
fact that a black hole coupled with charged matter fields may have
hairs and then will break the local U(1) gauge symmetry in the bulk. This happens when there is an instability towards a charged matter field to get an expectation value due to the bulk action and its boundary condition. To construct different kinds of holographic superconductor, we use different matter fields in the bulk action and solve the equation of motion to get the background. Generally speaking, this can be done only numerically. However, there always exists some parameter space which allows us to decouple the matter fields and the AdS black hole background. This decouple limit which we will use in the following sections is always taken place in the large charge limit of the matter fields and is called ``probe limit''.  Firstly, we introduce the AdS black hole background that will be used for several holographic superconductor models. The AdS black hole background can
be used only if it is a solution of the equations of motion. Then we will analytically compute the critical exponents for these models.

\subsection{The Anti-de Sitter black hole background}
According to the AdS/CFT dictionary, if we want to study a finite temperature field theory on the boundary, we need an AdS black hole background. The added matter fields are perturbation of the black hole in the probe limit and the metric will not be affected by
the added matter fields. The metric of the $AdS_4$ black hole in units in which the AdS radius $L= 1$ reads\cite{14}
\begin{equation}
ds^2=-f(r) dt^2+ r^2(dx^2+dy^2)+ \frac{dr^2}{f(r)},
\label{metric}
\end{equation}
where
\begin{equation}
f(r)=r^2-\frac{r_+^3}{r}.
\end{equation}
The Hawking temperature of the black hole is given by
\begin{equation}
T=\frac{3r_+}{4\pi},
\end{equation}
which is also the temperature of the dual gauge theory living on the
boundary of the AdS spacetime. This is a $AdS_4/CFT_3$ correspondence, which means that the
dual superconductor is of 2+1 dimension. In the probe limit, holographic superconductors
are constructed by adding perturbative matter fields to the black hole background.
For example, for the $p$-wave model we need $SU(2)$ gauge fields in the bulk.

\subsection{the $s$-wave model}
In this section we review the first analytic computation of
the critical exponent of the $s$-wave model which has been made in \cite{14}.
Other analytical methods have been developed in\cite{33,26} .
For review of the $s$-wave model one can see \cite{15,16,17,18}.
The matter fields of the $s$-wave model include a charged scalar field and a Maxwell field coupled to it. The condensation of the scalar operator on the boundary which corresponds to the scalar field in the bulk is the order parameter of the superconducting phase.

\begin{equation}
\mathcal{L}_s=-\frac{1}{4} F_{\mu \nu}F^{\mu \nu}-|\partial \Psi-iA \Psi|^2- m^2 |\Psi|^2
\end{equation}

We take the ansatz that $\Psi(r) \neq 0$ and $\Phi(r) \neq 0$, $\Phi(r)$ is the scalar potential \cite{6}. We introduce a new coordinate \be z = \frac{r_+}{r} \ee.

The equations of motion (EOMs) in the coordinate $z$ are
\begin{equation}
z\Psi'' - \frac{2+z^3}{1-z^3} \Psi' + \left[ z \frac{\Phi^2}{r_+^2 (1-z^3)^2} - \frac{m^2}{z(1-z^3)} \right] \Psi = 0
\end{equation}

\begin{equation}
\label{eq1}
\Phi'' - \frac{2\Psi^2}{z^2 (1-z^3)} \Phi = 0
\end{equation}
where the prime denotes differentiation with respect to $z$.
We have to solve the EOMs in the interval $(0,1)$, where $z=1$ is the horizon and $z=0$ is the boundary. As will be seen below, by solving the equation of $\Psi$, one can
obtain the value of $T_c$, while by solving the equation for $\Phi$,
one can find the critical behavior of the order parameter at $T_c$.
Near the boundary ($z\to 0$), the boundary conditions are
%\be
%z^2 \psi'' - 2z\psi' -m^2 \psi = 0 \ \ , \ \ \ \ \phi'' = 0 \ee
%showing that
\be  \label{eq100}
\Psi \approx \frac{\langle \mathcal{O}_{\Delta_\pm} \rangle }{\sqrt 2 r_+^{\Delta_\pm}} z^{\Delta_\pm} \ \ , \ \ \ \ \Phi  \approx \mu - \frac{\rho}{r_+} z \ee
where
\be \label{dimension} \Delta_\pm = \frac{3}{2} \pm \sqrt{\frac{9}{4} + m^2} \ee
$\langle \mathcal{O}_{\Delta_\pm} \rangle$ are the condensation with dimension $\Delta_\pm$, $\mu$ is the chemical potential and $\rho$ is the charge density of the field theory.

At the horizon, we require that the scalar field be regular and
\be\label{eq7} \Phi (1) = 0 \ee to ensure $A_t=\Phi dt$ to be well defined.

Since there is a second order continuous phase transition
at the critical temperature, the solution of the EOMs at $T_c$ is
\be \Phi (z) = \lambda r_{+c} (1-z) \ \ , \ \ \ \ \lambda = \frac{\rho}{r_{+c}^2} \ee
where $r_{+c}$ is the radius of the horizon at $T=T_c$.

As $T\to T_c$, the field equation for the scalar field $\Psi$ approaches the limit
\be - \Psi'' + \frac{2+z^3}{z(1-z^3)} \Psi' + \frac{m^2 }{z^2(1-z^3)} \Psi = \frac{\lambda^2 }{(1+z+z^2)^2} \Psi \ee

By solving the equation of $\Psi$, we can obtain the value of $T_c$.
To match the behavior at the boundary, we can define
\be\label{eq2} \Psi (z) = \frac{\langle \mathcal{O}_\Delta \rangle}{\sqrt 2 r_+^\Delta} z^\Delta F(z) \ee
where, according to eq.(\ref{eq100}), $F$ is normalized as $F(0)=1$.

We deduce
\be \label{eq55b1} - F'' + \frac{1}{z} \left[ \frac{2+z^3}{1-z^3} -2\Delta \right] F' + \frac{m^2 + 3 \Delta - \Delta^2 + \Delta^2 z^3 }{z^2(1-z^3)} F = \frac{\lambda^2 }{(1+z+z^2)^2} F \ee
when $z\to 0$,$\frac{F'}{z}$ should be finite, so this equation is to be solved subject to the boundary condition
\be F' (0) = 0 \ee

In order to compare the analytical results to the numerical results in \cite{6}, we set $m^2=-2$, then from eq.(\ref{dimension}) the
dimension $\Delta$ of the order parameter $\langle \mathcal{O}_\Delta \rangle$ are $\Delta=1$ or $\Delta=2$.
By putting  $m^2=-2$, the equation for $F$ reads

\be \label{eq55b} - F'' + \frac{1}{z} \left[ \frac{2+z^3}{1-z^3} -2\Delta \right] F' + \frac{\Delta^2 z }{1-z^3} F = \frac{\lambda^2 }{(1+z+z^2)^2} F \ee

Now, eq.(\ref{eq55b}) is translated to the Sturm-Liouwille form under the certain boundary condition. According to the general variation method to solve the Sturm-Liouwille problem (\cite{33},or see Appendix), the eigenvalue $\lambda^2$ minimizes the expression
\be\label{eq5a} \lambda^2 = \frac{\int_0^1 dz\ z^{2\Delta -2} \{ (1-z^3) [F'(z)]^2 + \Delta^2 z [F(z)]^2 \} }{\int_0^1 dz \ z^{2\Delta -2} \frac{1-z}{1+z+z^2} [F(z)]^2} \ee
In order to use the variation method , we have to specifies the trial eigenfunction $F(z)$. From the boundary condition $F(0)=1$ and $F'(0)=0$, the constant term is $1$ and the linear term of $z$ is forbidden. The second order trial eigenfunction is then
\be\label{eq5} F= F_\alpha (z) \equiv 1 - \alpha z^2 .\ee
If the lowest solution of $F(z)$ corresponds to $\alpha>0$, the whole solution of $\Phi(z)$  will decrease monotonically from the boundary to the horizon and it is believed to describe the superconductor vacuum solution \cite{18}. It will be seen in the following that this is true.

For $\Delta =1$ we obtain
\be \lambda_\alpha^2 = \frac{6-6\alpha + 10\alpha^2}{2\sqrt 3 \pi - 6\ln 3
+ 4 (\sqrt 3 \pi +3\ln 3 - 9)\alpha + (12\ln 3 - 13)\alpha^2} \ee
which attains its minimum at $\alpha \approx 0.239$. We obtain
\be \lambda^2 \approx \lambda_{0.239}^2 \approx 1.268 \ee
which can be compared with the numerical value $\lambda^2 = 1.245$ \cite{6}.
The critical temperature is
\be\label{eqTc} T_c = \frac{3}{4\pi} r_{+c} = \frac{3}{4\pi} \sqrt{\frac{\rho}{\lambda}} \ee
so for $\Delta = 1$, $T_c \approx 0.2250\sqrt\rho$, which is in very good agreement with the numerical value $T_c = 0.226\sqrt\rho$ \cite{6}.

In fact, this analytical calculation can be done even better if we include higher order of $z$ such as third order trial eigenfunction
$F= F_{\alpha \beta} (z) \equiv 1 - \alpha z^2 + \beta z^3 $. Then the eigenvalue $\lambda^2$ minimize eq.(\ref{eq5a}) when $\alpha \to 0.4962 $, $\beta \to 0.2706$ . As a result $\lambda^2 \approx 1.259$ and $T_c \approx 0.2255\sqrt\rho$  , which is more close to the numerical result. However, for qualitative analyze, the second order trial eigenfunction is good enough and we use it in the following sections.

Similarly, for $\Delta=2$ we obtain
\be \lambda_\alpha^2 = 2 \frac{1 - \frac{4}{3}\alpha + \frac{4}{5} \alpha^2}{3-\ln 3 - \frac{\pi}{\sqrt 3} + (\frac{13}{3} - 4\ln 3)\alpha + ( \frac{\pi}{\sqrt 3} - \frac{7}{10} + \ln 3)\alpha^2} \ee
whose minimum is
$\lambda^2 \approx 17.3$ (at $\alpha\approx 0.6$), which can be compared with the exact value $\lambda^2 = 16.754$.
The critical temperature in this case is $T_c \approx 0.117\sqrt\rho$, which is in very good agreement with the numerical value $T_c = 0.118\sqrt\rho$ \cite{6}.

Now we begin to solve the equation for $\Phi$ to obtain the
behavior of the order parameter at $T_c$. Away from (but close to) the critical temperature, the field eq.(\ref{eq1}) for $\Phi$ becomes
\be
\label{eq4}
\Phi'' = \frac{\langle\mathcal{O}_\Delta\rangle^2}{r_+^{2\Delta}}\ \frac{z^{2(\Delta -1)} F^2(z)}{1-z^3}\ \Phi \ee
where the parameter $\langle\mathcal{O}_\Delta\rangle^2/(2r_+^{2\Delta})$ is small.
We may expand $\Phi$ in the small parameter as
\be \frac{\Phi}{r_+} = \lambda (1-z) + \frac{\langle\mathcal{O}_\Delta\rangle^2}{r_+^{2\Delta}} \chi (z) + \dots \ee
where $\chi$ is the general correction function. For the correction $\chi$ near the critical temperature, from eq.(\ref{eq4})we deduce
\be
\chi'' = \lambda \frac{z^{2(\Delta -1)} F^2(z)}{1+z+z^2} \ee
with $\chi(1) = \chi'(1)=0$.

To find the temperature, we need
\be\label{eq6} \chi' (0) = \lambda \mathcal{C} \ \ , \ \ \ \
\mathcal{C} = \int_0^1 dz \ \frac{z^{2(\Delta -1)} F^2(z)}{1+z+z^2} \ee
From eq.(\ref{eq100}), we deduce the ratio
\be \label{eq8}
\frac{\rho}{r_+^2} = \lambda \left( 1 +  \frac{\mathcal{C} \langle\mathcal{O}_\Delta\rangle^2}{r_+^{2\Delta}} + \dots \right) \ee
Using the fact that $T_c =3 \sqrt{\rho/\lambda}/4\pi$ and $T=3r_{+}/4\pi$, the equation becomes
\begin{equation}
T_c^2-T^2 \propto \frac{T^2\langle\mathcal{O}_\Delta\rangle^2}{T^{2\Delta}}
\end{equation}
Then
\begin{equation}
T^{2\Delta-2}(T_c+T)T_c(1-\frac{T}{T_c})\propto \langle\mathcal{O}_\Delta\rangle^2.
\end{equation}
Since $T$ is very close to $T_c$, we have
\be \label{eqgam}
\langle \mathcal{O}_\Delta \rangle \propto
T_c^\Delta \left( 1 - \frac{T}{T_c} \right)^{1/2} \ \ , \ \ \ \
\ee

From the computations from eq.(\ref{eq4}) to eq.(\ref{eqgam}) we can see that the critical exponent
$1/2$ comes from two aspects: the first one is $T_c \propto \sqrt{\rho/\lambda}$, the second one is
the $\Psi^2 \Phi$ term in eq.(\ref{eq4}) which determines the order of the order parameter in eq.(\ref{eq8})
to be 2. The fact that $T_c \propto \sqrt{\rho/\lambda}$ is from dimensional
analysis, while the second order of $\langle \mathcal{O}_\Delta \rangle$
in eq.(\ref{eq4}) is determined by the $\Psi^2 \Phi^2$ term in $\mathcal{L}_s$, and this
term is determined by the fundamental symmetry. In Ref.\cite{30}
the author also found the results that the critical exponent $1/2$ follows from the $\Psi^2 \Phi$ coupling.
The fundamental
symmetry includes the local $U(1)$ gauge symmetry and the positive/negative symmetry of the order parameter.
The latter one determines that there are only quadratic and quartic terms in the action. As we will see in section 3,
the critical exponent differs from $1/2$ when the added cubic term breaks the positive/negative symmetry of the order parameter.
If one adds quartic terms in the action, the value of the critical exponent will not be altered.
The cubic term is not allowed in the G-L theory, since we have
to preserve the positive/negative symmetry of the order parameter as a fundamental
symmetry. If we add the cubic term in the G-L theory anyway, we will naturally
get a critical exponent different from $1/2$.
If we put the ansatz $\Psi(r) \neq 0$ and $\Phi(r) \neq 0$ into the action
and do the computation, we will get
\begin{equation}
\mathcal{L}_s=-\frac{1}{f}\Phi^2 \Psi^2 + f (\partial_z \Psi)^2 + m^2 \Psi^2 -(\partial_z \Phi)^2
\end{equation}
This Lagrangian has the same symmetry as the G-L Lagrangian.
\subsection{the $p$-wave model}
The matter field of the $p$-wave holographic superconductor is a pure
SU(2) gauge field \cite{7,8,9}.
\begin{eqnarray}
\mathcal{L}_{\textmd{p}}=-\frac{1}{2g_{\rm YM}^2}\textmd{Tr}(F_{\mu\nu}F^{\mu\nu}),
\end{eqnarray}
where $g_{\rm YM}$ is the gauge coupling constant and
$F_{\mu\nu}=T^aF^a_{\mu\nu}=\partial_\mu A_\nu-\partial_\nu
A_\mu-i[A_\mu,A_\nu]$ is the field strength of the gauge field
$A=A_\mu dx^\mu=T^aA^a_\mu dx^\mu$. For the $SU(2)$ gauge symmetry,
$[T^a,T^b]=i\epsilon^{abc}T^c$ and
$\textmd{Tr}(T^aT^b)=\delta^{ab}/2$, where $\epsilon^{abc}$ is the
totally antisymmetric tensor with $\epsilon^{123}=1$. The Yang-Mills
Lagrangian becomes
$\textmd{Tr}(F_{\mu\nu}F^{\mu\nu})=F^a_{\mu\nu}F^{a\mu\nu}/2$ with
the field strength components $F^a_{\mu\nu}=\partial_\mu
A^a_\nu-\partial_\nu A^a_\mu+\epsilon^{abc}A^b_\mu A^c_\nu$.

The ansatz for the $p$-wave background is $A=\phi(z)T^3dt + w(z)T^1dx.$ with other components vanishing.
Here the $U(1)$ subgroup of $SU(2)$ generated by $T^3$ is identified
to be the electromagnetic gauge group \cite{7} and $\phi$ is
the electrostatic potential, which must vanish at the horizon for
the gauge field $A$ to be well-defined, but need not vanish at
infinity. Thus the black hole can carry charge through the
condensate $w$, which spontaneously breaks the $U(1)$ gauge
symmetry.
The EOMs for the two fields $\phi$ and $w$ are
\begin{equation}
z w''-\frac{3z^3}{1-z^3} w'+ \frac{z\phi^2}{r_+^2(1-z^3)^2} w=0
\end{equation}

\begin{equation}
\phi''-\frac{w^2}{1-z^3}\phi=0
\end{equation}

It is clear that the coupled term of the condensed matter field and the
scalar potential is $w^2 \phi$, in which the condensed matter field $w$ is
quadratic. This is similar to the $s$-wave model which is
important to obtain the critical exponent $1/2$, as we have discussed
in the last section.
Now our task is to solve the EOMs in the interval (0,1) under
the proper boundary conditions. The boundary condition at $z=0$
is
\begin{equation}
w\approx\frac{\langle \mathcal{O}\rangle }{r_+}z, \ \ \ \  \ \ \phi\approx\mu-\frac{\rho}{r_+}z.
\end{equation}
$\mathcal{O}$ is the order parameter, $\mu$ is the chemical potential and
$\rho$ is the charge density of the field theory. In the $s$-wave and $d$-wave
model, the dimension of the order parameter is determined by the value of the mass.
While in this $p$-wave model the order parameter is just of dimension one.
At the horizon, we require $w$ be regular and $\phi=0$, since $\phi dt$ needs to be well
defined at $z=1$.

If there is a second order continuous phase transition at the critical
temperature, the solution of the EOMs at the $T_c$ should be
\begin{equation}
w(z)=0, \phi(z)=\lambda r_{+c}(1-z),
\end{equation}
where $\lambda=\frac{\rho}{r_{+c}^2}$, $r_{+c}$ is the radius of
horizon corresponding to $T=T_c$.
So, at a temperature slightly below $T_c$, the EOM for
$w$ becomes:
\begin{equation}
-w''+\frac{3 z^3}{z(1-z^3)}w'=\frac{\lambda^2}{(1+z+z^2)^2}w.
\end{equation}
We can define $w$ as
\begin{equation}
w(z)=\frac{\langle \mathcal{O}\rangle }{r_+} z F(z)
\end{equation}
To math the boundary condition at $z=0$, $F(0)$ should be normalized
as $F(0)=1$. Then we can get the equation for $F(z)$,
\begin{equation}
-F''+\frac{1}{z}(\frac{3 z^3}{1-z^3}-2)F'+\frac{3 z F(z)}{(1-z)^3}=\frac{\lambda^2}{(1+z+z^2)^2}F
\end{equation}
By solving this equation, we can obtain the value of the critical temperature. Besides the boundary condition $F(0)=1$,
another boundary condition is $F'(0)=0$. Then we can solve this
equation by using the same method for the $s$-wave model.
The eigenvalue $\lambda$ minimizes the expression
\begin{equation}
\lambda^2 = \frac{\int_0^1 dz\ z^2\{(1-z^3) [F'(z)]^2 + 3z [F(z)]^2 \} }{\int_0^1 dz \ z^2 \frac{1-z}{1+z+z^2} [F(z)]^2}
\end{equation}
To estimate it, we use the trial function
\begin{equation}
 F= F_\alpha (z) \equiv 1 - \alpha z^2
\end{equation}
We then obtain
\begin{equation}
\lambda_\alpha^2 = \frac{3-\alpha + \frac{27}{10}\alpha^2}{6-2\ln 3-\frac{2\sqrt3}{3}\pi-(8\ln3-\frac{26}{3})\alpha-(\frac{7}{5}+2\ln3-\frac{2\sqrt{3}}{3} \pi)\alpha^2}
\end{equation}
which attains its minimum at $\alpha\approx 0.51$. We obtain
\begin{equation}
\lambda^2\approx\lambda_{0.51}^2\approx 13.77
\end{equation}
The critical temperature is
\begin{equation}
T_c = \frac{3}{4\pi} r_{+c} = \frac{3}{4\pi} \sqrt{\frac{\rho}{\lambda}}
\end{equation}
So, $T_c\approx 0.124\sqrt\rho$, which is in agreement with the numerical result $T_c\approx 0.125\sqrt\rho$ in \cite{9}.

If we want to know the behavior of the order parameter at $T_c$, we need to solve
the equation for the scalar potential close to $T_c$,
\begin{equation}
\phi''=\frac{\langle \mathcal{O}\rangle^2}{r_+^2}\frac{z^2 F^2(z)}{1-z^3}\phi.
\end{equation}
Since the order parameter $\langle \mathcal{O}\rangle$ is small, $\frac{\langle \mathcal{O}\rangle^2}{r_+^2}$
is a small parameter. We can expand $\phi$ in this small parameter as
\begin{equation}
\frac{\phi}{r_+}=\lambda(1-z)+\frac{\langle \mathcal{O}\rangle^2}{r_+^2}\chi(z)+\cdots.
\end{equation}
Then we get the equation for $\chi$,
\begin{equation}
\chi''=\lambda\frac{z F^2(z)}{1+z+z^2},
\end{equation}
and $\chi(1)=\chi'(1)=0$.

From the asymptotic behavior of $\phi$ at the boundary, we have $\phi'(0)=\rho/r_+$.
We can also get $\chi'(0)$ from the above equation,
\begin{equation}
\chi'(0)=\lambda C, C=\int_0^1 dz \frac{z^2 F^2(z)}{1+z+z^2}.
\end{equation}
 We have
\begin{equation}
\frac{\rho}{r_+^2}=\lambda(1+\frac{C\langle \mathcal{O}\rangle^2}{r_+^2}+\cdots).
\end{equation}
Then
\begin{equation}
\langle \mathcal{O}\rangle \propto  T_c(1-\frac{T}{T_c})^{1/2},
\end{equation}
which is the exact mean field critical exponents given by numerical calculations.
The $p$-wave model is different from the $s$-wave and $d$-wave model.
In this $p$-wave model the condensed charged matter field is not coupled to the
$U(1)$ via the covariant derivative. It is hard to reveal the fundamental
symmetry of the $p$-wave action. If one puts the ansatz into the action, it is clear
that the action has the same fundamental symmetry as the $s$-wave model
with terms like $(\partial_z w)^2$, $ (\partial_z \phi)^2$ and $w^2 \phi^2$.

\subsection{the $d$-wave model}
In order to construct a holographic model of d-wave superconductor, we need
a spin two field in the bulk \cite{10}.
In \cite{11,12}, a d-wave model with charged
tensor field and Maxwell field in the bulk is built, in which the inconsistency
of charged tensor field in a curved background like
the appearance of ghost and causality can be made very small in some limit. The matter field
of the holographic d-wave model is,
\begin{equation}
\begin{split}
\label{E:ActionSimp}
\mathcal{L}_d &= - |D_\rho \varphi_{\mu\nu}|^2 + 2|D_\mu \varphi^{\mu\nu}|^2 + |D_\mu \varphi|^2 - \big[ D_\mu \varphi^{*\mu\nu} D_\nu \varphi + \text{c.c.} \big] - m^2 \big( |\varphi_{\mu\nu}|^2 - |\varphi|^2 \big) \\
&\quad +2  R_{\mu\nu\rho\lambda} \varphi^{*\mu\rho} \varphi^{\nu\lambda}
- R_{\mu\nu} \varphi^{*\mu\lambda} \varphi^\nu_\lambda
- \frac{1}{4} R | \varphi |^2
- i  q F_{\mu\nu} \varphi^{*\mu\lambda} \varphi^\nu_\lambda - \frac14 F_{\mu\nu} F^{\mu\nu} \;,
\end{split}
\end{equation}
where $D_\mu = \nabla_\mu - i q A_\mu$ and $\varphi_\rho = D^\mu \varphi_{\mu\rho}$.
The equations of motion which follow from (\ref{E:ActionSimp}) are
\bea
\label{EOM probe action}
& 0 = (\square - m^2) \varphi_{\mu\nu} - 2 D_{(\mu} \varphi_{\nu)} + D_{(\mu} D_{\nu)} \varphi - g_{\mu\nu} \big[ (\square - m^2) \varphi - D^\rho \varphi_\rho \big] \\
&\qquad + 2 R_{\mu\rho\nu\lambda} \varphi^{\rho\lambda} - g_{\mu\nu} \frac {R}{d+1} \varphi - i\frac q2 \big( F_{\mu\rho} \varphi^\rho_\nu + F_{\nu\rho} \varphi^\rho_\mu \big) \\
& D_\mu F^{\mu\nu} = J^\nu
\eea
where
\be
J^\nu = i \varphi^*_{\alpha\beta} (D^\nu \varphi^{\alpha\beta} - D^\alpha \varphi^{\nu\beta}) + i(\varphi^*_\alpha - D_\alpha \varphi^*)(\varphi^{\nu\alpha} - g^{\nu\alpha} \varphi) + \text{h.c.} \;.
\ee

For the $d$-wave backgrounds, the ansatz takes the following form \cite{11,12}
\begin{equation}
\label{ansatz}
A = A_\mu \, dx^\mu \equiv  \phi(z) \, dt  \;, \qquad\qquad
\psi_{xy}(z) \equiv \frac{L^2}{2z^2} \, \psi(z) \;,
\end{equation}
with all other components of $\psi_{\mu\nu}$ set to zero, and $\phi$ and $\psi$ are real. The ansatz \eqref{ansatz} satisfies $\psi = \psi_\mu = F_{\mu\rho} \psi^\rho_\nu = 0$. Instead of turning on $\psi_{xy}$ in \eqref{ansatz} we could consider a non-vanishing value for $\psi_{xx-yy} \equiv \psi_{xx} = - \psi_{yy}$. These two ansatzs are equivalent under a
$\pi/4$ rotation \cite{12}.

With this ansatz, we can derive the equations of motion for the two
fields $\psi$ and $\phi$,

\begin{equation}
z\psi'' - \frac{2+z^3}{1-z^3} \psi' + \left[ z \frac{\phi^2}{r_+^2 (1-z^3)^2} - \frac{m^2}{z(1-z^3)} \right] \psi = 0
\end{equation}

\begin{equation}
\phi'' - \frac{\psi^2}{z^2 (1-z^3)} \phi = 0
\end{equation}

It is interesting that the EOMs for the condensed field and scalar potential are the same as the ones for the $s$-wave model except for the coefficient of the term $\psi^2 \phi$ in the equation for $\phi$. But the boundary conditions at the boundary are slightly different,

\be\label{eq3} \psi \thickapprox z^{\Delta} \frac{\langle \cO_{xy} \rangle}{r_+^{\Delta}( 2\Delta - 3)} \ \ , \ \ \ \ \phi \thickapprox \mu - \frac{\rho}{r_+} z \ee
where $\rho$ is the charge density of the boundary field theory, $\Delta$ is given by $m^2 L^2 = \Delta(\Delta - 3)$, and $m^2\geq0$ \cite{12}.
We can check that such a choice of dimension is compatible with the unitarity bounds in conformal theories, for instance \cite{22}.
In addition, in Ref. \cite{12}, the authors give a generalization of the analysis of Breitenlohner and Freedman \cite{23} to obtain the bound of $m^2\geq0$.
Since the EOMs of the $d$-wave model are the same to those of the $s$-wave model and the boundary condition with different $\Delta$ will not affect our results of the critical exponents, we can conclude that for the $d$-wave model, the order parameter behaves as

\begin{equation}
\langle \cO_{xy} \rangle \propto T_c^\Delta(1-\frac{T}{T_c})^{1/2}
\end{equation}

\section{The extended $s$-wave model}
In order to construct a holographic superconductor with
 critical exponents different from $1/2$, we need to
extend the $s$-wave model by preserving the gauge symmetry. First
we can rewrite the $s$-wave model innocuously in a St\"{u}ckelberg form by
rewriting the charged scalar field as $\Psi e^{ip}$:

\begin{equation}
\mathcal{L}_s=-\frac{1}{4} F_{\mu \nu}F^{\mu \nu}-\partial \Psi^2-\Psi^2(\partial p-A)^2 - m^2 |\Psi|^2
\end{equation}
So far all we have done is to rewrite the model. Nevertheless, it is straightforward to generalize the model in a gauge invariant way. The generalized action reads

\begin{equation}
\mathcal{L}_s=-\frac{1}{4} F_{\mu \nu}F^{\mu \nu}-\partial \Psi^2-|\mathcal{K}(\Psi)|(\partial p-A)^2 - m^2 |\Psi|^2
\end{equation}
This is called the St\"{u}ckelberg Lagrangian\cite{24,25,26}. The general form of $\mathcal{K}(\Psi)$ is
\begin{equation}
\mathcal{K}(\Psi)=|\Psi^2| + c_3 |\Psi|^3 + c_4 |\Psi|^4,
\end{equation}
In the above equation we have taken the absolute value since we require the Lagrangian still be
local U(1) gauge invariant. In this model, the probe limit also works in the large charge limit\cite{25}.
When $c_3$ and $c_4$ vanish, the model reduces to the original $s$-wave
model. When $c_3=0$ and $0 <c_4< 1.4$ \cite{24}, the model has a second order phase transition with a critical exponent $1/2$ at the critical temperature. Since this is still
a continuous phase transition, we can apply the analytical analysis, as will be
discussed in detail below. When $c_4>1.4$, the superconducting
phase transition turns to be of first order and our analytic method breaks down.

With the ansatz $\Phi(z)\neq 0 , \Psi \neq 0$ and the gauge freedom to fix $p = 0$. we have the EOMs:
\bea
z\Psi'' - \frac{2+z^3}{1-z^3} \Psi' +  z \frac{\Phi^2}{2 r_+^2 (1-z^3)^2}\mathcal{K'} - \frac{m^2}{z(1-z^3)}\Psi &=& 0 \nonumber\\
\Phi'' - \frac{2\mathcal{K}}{z^2 (1-z^3)} \Phi &=& 0
\eea
The boundary condition for $\Psi$ and $\Phi$ are the same as that of the $s$-wave model.
It is straightforward to repeat the computation for $s$-wave model here with also $m^2=-2$. Now,
the equation for $F$ becomes
\be\label{eq5b} - F'' + \frac{1}{z} \left[ \frac{2+z^3}{1-z^3} -2\Delta \right] F' + \frac{\Delta^2 z }{1-z^3} F = \frac{\lambda^2 }{(1+z+z^2)^2} (F+2c_4F^3) \ee

Since $F$ is a small near the critical point, it is reasonable to ignore the $2c_4F^3$ term if we are only concerned about the behavior near the critical point. After solving the equation of $F$ with the same method used before, we obtain $T_c \approx 0.117\sqrt\rho$ for $\Delta = 2$. For different values of $c_4$, the critical temperature is the same. This conclusion is in agreement with the numerical computation in Ref. \cite{24}.

We now solve the equation for $\Phi$ to see if the mean field behavior will be altered by
$\mathcal{K}$. The equation for $\Phi$ close to $T_c$ is
\begin{equation}
\Phi'' = \big( \frac{\langle\mathcal{O}_\Delta\rangle^2}{r_+^{2\Delta}}\ \frac{z^{2\Delta } F^2(z)}{z^2(1-z^3)}+ \frac{c_4\langle\mathcal{O}_\Delta\rangle^4}{2r_+^{4\Delta}}\ \frac{z^{4\Delta } F^4(z)}{z^2(1-z^3)} \big)\Phi,
\end{equation}
where $c_4 < 1.4$. Similar to the computation in the above sections, we expand $\Phi$ in the small parameter $\frac{\langle\mathcal{O}_\Delta\rangle^2}{r_+^{2\Delta}}$ as
\begin{equation}
\frac{\Phi}{r_+} = \lambda (1-z) + \frac{\langle\mathcal{O}_\Delta\rangle^2}{r_+^{2\Delta}} \chi (z) + \dots.
\end{equation}
 We can get the equation for $\chi$
\begin{equation}
\chi''=\frac{\lambda}{z^2(1+z+z^2)}\big( F^2 z^{2\Delta}+\frac{\langle\mathcal{O}_\Delta\rangle^2}{2r_+^2} z^{4\Delta} F^4 \big).
\end{equation}

In order to match the asymptotic behavior of $\Phi$ at the boundary, we have
\begin{equation}
\label{eq9}
\frac{\rho}{r_+^2}=\lambda(1+\frac{\mathcal{C}_1\langle\mathcal{O}_\Delta\rangle^2}{r_+^{2\Delta}}+\frac{\mathcal{C}_2\langle\mathcal{O}_\Delta\rangle^4}{2r_+^{4\Delta}}+\cdots),
\end{equation}
where
\begin{equation}
\mathcal{C}_1= \int_0^1 dz \ \frac{z^{2\Delta } F^2(z)}{z^2(1+z+z^2)}  , \mathcal{C}_2= \int_0^1 dz \ \frac{c_4 z^{4\Delta } F^4(z)}{z^2(1+z+z^2)}
\end{equation}
Solving eq.(\ref{eq9}) for $\langle\mathcal{O}_\Delta\rangle$ and selecting the only
physical solution with $\langle\mathcal{O}_\Delta\rangle > 0$, we get
\begin{equation}
\langle\mathcal{O}_\Delta\rangle \propto T_c^\Delta(1-T/T_c)^{1/2}.
\end{equation}
This is in agreement with the numerical results in \cite{24}.

  The critical behavior with critical exponent $1$ appears when the $|\Psi|^3$ does not vanish. The situation
with $c_3=-1$ and $c_4=0.4$ was studied numerically in \cite{24,25} with the
critical exponent being 1 rather than 1/2. The order parameter has both
positive and negative solutions since the cubic term breaks the symmetry
between positive and negative $\Psi$, see Fig. (2) in \cite{24}. Since a negative order parameter is unphysical, we must get rid of the solution with
negative values. The analytical method we use here is based on the
fact that when we lower the temperature from the normal state with the
condensation vanishing to $T_c$, the order parameter goes continuously from zero
to a finite value at the critical temperature. While for the situation
with $c_3=-1$, numerical calculations tell us that if we lower the temperature from
a high value, we get a negative order parameter before we get the
finite positive physical order parameter. So the analytical method we
use here are not feasible now. However, just as was shown in Ref. \cite{26}, another analytic method is
developed to find the relationship between the value of critical exponents and $\mathcal{K}$.
The cubic term breaks the positive/negative symmetry of the order parameter. This
is not allowed in the G-L theory, since we have only squared and quartic terms which
preserve the positive/negative symmetry of the order parameter.

Since the mean-field behavior is protected by the fundamental
symmetry of the G-L theory, even though different from the original $s$-wave model, the extended $s$-wave model contains a quartic term, we still have the same mean field critical behaviors. Therefore, if we add a cubic term in the $s$-wave model to break this symmetry, the appearance of
so called ``non-mean field''  critical exponent which differs from the usual value of $1/2$ can be understandable. However, the model with non-vanish $c_3$ has problem when we investigate the target space of the model. In (3.1) the action of a complex scalar is rewritten in terms of real
fields $\Psi$(its modulus) and $p$ (its phase). In (3.2) the kinetic term is modified:
the action is still gauge-invariant, but it describes a $\sigma$-model whose target
space is parametrized by $\varphi^i = (\Psi	, p)$ and it is in general not flat. The kinetic
term of the $\sigma$ model is
\begin{equation}
-g_{ij}\partial_{\mu}\varphi^i\partial^{\mu}\varphi^j.
\end{equation}
Using the expression (3.3) the metric is
\begin{equation}
g_{ij} = \textrm{diag}(1,\Psi^2 + c_3\Psi^3 + c_4\Psi^4).
\end{equation}

When $c_3 = c_4 = 0$, it is the first holographic superconductor
in \cite{6}. It is clear that the metric is flat, so the $\sigma$-model just
describes a complex scalar field.
For the case of  $c_4 \neq 0$,  $c_3 = 0$, the target space is no longer flat: its scalar
curvature is
\begin{equation}
R =-\frac{2c_4(3+2c_4\Psi^2)}{(1+c_4\Psi^2)^2}
\end{equation}
 The target
space is curved, but it is still a smooth two-dimensional manifold with no singularities,
so the model makes perfect sense.
However, the story is different when turning on $c_3$ (and setting $c_4 = 0$, as turning on $c_4$ as well does not change
the conclusion) the scalar curvature is
\begin{equation}
R =-\frac{c_3(4+3c_3\Psi)}{2\Psi(1+c_3\Psi)^2} .
\end{equation}
When $\Psi\rightarrow0$ the scalar curvature diverges. Therefore the target
manifold has a singularity at the origin.
In the classical theory, we could make sense of the $\sigma$-model by removing the
singular point. However in the holographic model this looks quite awkward,
because above the critical temperature the classical solution is precisely $\Psi=
0$, that is it sits at the singular point. This indicates that the theory with $c_3\neq 0$
have problem.\footnote{The discussion of the
feasibility of the extend $s$-wave model is based directly on the referee's argument.
We thank the referee for pointing this important fact.} However, we worked in the probe limit, the theory may make sense
when we include the back reaction of matter field on the metric of the AdS black hole.

\section{Discussion}
  In this paper, under the large $N$ limit when the
quantum fluctuation is suppressed, we analytically study the
mean field behaviors of the order parameter at
$T_c$ for four different holographic models of superconductor and it is found that these four models have similar properties. For each
model there is a charged matter field coupled to a
$U(1)$ background gauge field. After solving the two non-linear coupled EOMs
with proper boundary conditions, the information we need
for the strongly coupled boundary theory can be obtained from the asymptotic behaviors
of the matter field and gauge field near the boundary.
Numerically it is hard to see what properties of the bulk theory leads to the
mean field or ``non-mean-field'' behaviors. However, by using the analytical method, it is seen
 that the equation of the charged condensed matter field gives the
value of $T_c$ while the equation of the scalar potential gives the
behavior of order parameter at $T_c$. If the bulk theory has the same fundamental
symmetry as the G-L theory, then the equation for the scalar potential leads to the
mean field critical exponent.
When the added cubic term
breaks the symmetry, the critical exponent differs from $1/2$. The so-called ``non mean field'' behavior in Ref. \cite{24} is just the result of violation of the symmetry of the usual mean field theory like the G-L theory.
In the present paper, we focus on the $AdS_4/CFT_3$ situation, in which
the dual superconductor is of $2+1$ dimension.
The universal critical exponents of $s$-wave holographic superconductors in various spacetime dimensions have been studied in \cite{31}, in which the authors found that in the large $N$ limit, the mean-field results are independent of dimension just like the G-L theory.
It would be interesting to compute $1/N$-effects to see
how the fluctuation will
affect the ``mean-field'' behavior in the holographic superconductors\cite{32}.
We obtain our conclusion by comparison of several holographic superconductors with mean field or non mean field critical exponents
and the results are convincing. However, it is a better way to apply the holographic renormalization
group \cite{27,28,29} analysis to these models to find the low energy
effective theories of holographic superconductors. This may bridge the gap between the holographic models of superconductor and the G-L theory.

\section*{Acknowledgements}

 We especially thank the referee for
 pointing out the problem of the extend $s$-wave model when $c_3>0$.
 We thank M. Kaminski for discussion on
the extended $s$-wave model. We also would like to thank Wei-Min Sun for valuable comments.
H.B. Zeng, Y. Jiang and H.S. Zong are supported in part by the NSFC ( Grant Nos. 10775069 and 10935001) and the Research Fund for the Doctoral Program of Higher Education (Grant No. 20080284020). X. Gao is supported in part by the NSFC (Grant No.10821504). This research was also supported in part by the Project of Knowledge Innovation Program (PKIP) of Chinese Academy of Sciences, Grant No. KJCX2.YW.W10.

\appendix
\renewcommand{\theequation}{\thesection.\arabic{equation}}
\addcontentsline{toc}{section}{Appendices}
\section*{APPENDIX}

\section{Variation method to solve the Sturm-Liouville problem}

To familiar the readers to the variation method to solve the Sturm-Liouville problem, we present some basic results of this method and all the derivations can be found in many text books such as \cite{33} . The Sturm-Liouville  eigenvalue problem is to solve the equation
\begin{equation}
\frac{d}{dx}[k(x)\frac{dy}{dx}]-q(x)y(x)+\lambda\rho(x)y(x)=0
\end{equation}
with boundary condition
\begin{equation}
k(x)y(x)y'(x)|_a^b=0.
\end{equation}
The Sturm-Liouville problem can be result to be a functional minimize problem:
\begin{equation}
\label{eq10} F[y(x)]=\frac{\int_a^b dx (k(x) y'(x)^2+q(x) y(x)^2)}{\int_a^b dx \rho(x) y(x)^2}
\end{equation}
where the minimal eigenvalue $\lambda_0$ and its eigenstate $y_0(x)$ can be obtained by variate the above equation.
The $n+1$ th eigensystem can also be obtained by variation eq(\ref{eq10}) with constrains below:
\begin{equation}
\int_a^b dx \rho(x) y^*_n(x) y_i(x) = \delta_{n,i}\ \ \ \ \ \ \ \ \ \ (i=0,1,...,n-1,n).
\end{equation}
then, the eigenvalue $\lambda_n$ satisfies:
\begin{equation}
\lambda_n\leq\frac{\int_a^b dx (k(x) y'(x)^2+q(x) y(x)^2)}{\int_a^b dx \rho(x) y(x)^2}
\end{equation}
with complete eigenfunction \{$y_n(x)$\}.

\end{document}